**Adaptive simulated annealing (ASA): Lessons learned**

Lester Ingber

Lester Ingber Research
P.O.B. 857
McLean, VA 22101
U.S.A.
ingber@alumni.caltech.edu

**ABSTRACT**

Adaptive simulated annealing (ASA) is a global optimization algorithm based on an associated proof that the parameter space can be sampled much more efficiently than by using other previous simulated annealing algorithms. The author's ASA code has been publicly available for over two years. During this time the author has volunteered to help people via e-mail, and the feedback obtained has been used to further develop the code. Some lessons learned, in particular some which are relevant to other simulated annealing algorithms, are described.

## 1. INTRODUCTION

Adaptive simulated annealing (ASA) is a global optimization algorithm that relies on randomly importance-sampling the parameter space, i.e., in contrast to utilizing deterministic approaches often used by OR and mathematical programming people.

Since the public release of the very fast simulated reannealing (VFSR) code (Ingber, 1989), now called ASA (Ingber, 1993a), I have volunteered to help users via e-mail on code-specific problems they might encounter. The popularity of the code, roughly measured by the size of the ASA_list of people specifically requesting to be placed on a monthly list of updates—now at about 500 names—increased about a factor of two after an article in *The Wall Street Journal* (Wofsey, 1993). I have had contact with thousands of users via e-mail, a few via other channels of communication, and it is clear that these are just a small fraction of people that have at least tried to use the code. Therefore, I conclude that the code is quite stable and free of major bugs.

This paper deals with some of the lessons learned from this interaction. I think that at least some of these lessons may be useful to other developers of simulated annealing (SA) code, as well as to many users. A previous paper has described SA in the context of its use across many disciplines, and the variation in use in actual practice versus the algorithm offered in theoretical papers (Ingber, 1993b).

Section 2 gives a brief mathematical description of the ASA algorithm. Section 3 describes some of the options in the code, and explains why they are useful. Section 4 addresses some comparison tests and controversial claims, emphasizing that in fact there are many shades of SA. Section 5 gives a brief conclusion.

ASA source code in C-language is publicly available via anonymous ftp (Ingber, 1993a), or via the world wide web (WWW) at http://www.alumni.caltech.edu/˜ingber/. Problems with the code can be addressed to ingber@alumni.caltech.edu. Requests to be placed on the ASA mailing list should be addressed to asa−request@alumni.caltech.edu.

## 2. ASA ALGORITHM

## 2.1. "Standard" Simulated Annealing (SA)

The Metropolis Monte Carlo integration algorithm (Metropolis *et al*, 1953) was generalized by the Kirkpatrick algorithm to include a temperature schedule for efficient searching (Kirkpatrick *et al*, 1983). A sufficiency proof was then shown to put an lower bound on that schedule as $1/\log(t)$, where $t$ is an artificial time measure of the annealing schedule (Geman and Geman, 1984). However, independent credit usually goes to several other authors for independently developing the algorithm that is now recognized as simulated annealing (Cerny, 1982; Pincus, 1970).

### 2.1.1. Boltzmann annealing (BA)

Credit for the first simulated annealing is generally given to a Monte Carlo importance-sampling technique for doing large-dimensional path integrals arising in statistical physics problems (Metropolis *et al*, 1953). This method was generalized to fitting non-convex cost-functions arising in a variety of problems, e.g., finding the optimal wiring for a densely wired computer chip (Kirkpatrick *et al*, 1983). The choices of probability distributions described in this section are generally specified as Boltzmann annealing (BA) (Szu and Hartley, 1987).

The method of simulated annealing consists of three functional relationships.

1. $g_T(x)$: Probability density of state-space of $D$ parameters $x = \{x^i; i = 1, D\}$, where the subscript $T$ signifies a parameterization popularly referred to as the "temperature."

2. $h(\Delta E)$: Probability for acceptance of new cost-function given the just previous value.

3. $T(k)$: schedule of "annealing" the "temperature" $T$ in annealing-time steps $k$, i.e., of changing the volatility or fluctuations of one or both of the two previous probability densities.

The acceptance probability is based on the chances of obtaining a new state with "energy" $E_{k+1}$ relative to a previous state with "energy" $E_k$,

$$h(\Delta E) = \frac{\exp(-E_{k+1}/T)}{\exp(-E_{k+1}/T) + \exp(-E_k/T)}$$

$$= \frac{1}{1 + \exp(\Delta E/T)}$$

$$\approx \exp(-\Delta E/T) , \tag{1}$$

where $\Delta E$ represents the "energy" difference between the present and previous values of the energies (considered here as cost functions) appropriate to the physical problem, i.e., $\Delta E = E_{k+1} - E_k$. This essentially is the Boltzmann distribution contributing to the statistical mechanical partition function of the system (Binder and Stauffer, 1985).

This sampling algorithm also can be described by considering: a set of states labeled by $x$, each with energy $e(x)$; a set of probability distributions $p(x)$; and the energy distribution per state $d(e(x))$, giving an aggregate energy $E$,

$$\sum_x p(x)d(e(x)) = E \ . \tag{2}$$

The principle of maximizing the entropy, $S$,

$$S = -\sum_x p(x) \ln[p(x)/p(\bar{x})] \ , \tag{3}$$

where $\bar{x}$ represents a reference state, using Lagrange multipliers (Mathews and Walker, 1970) to constrain the energy to average value $T$, leads to the most likely Gibbs distribution $G(x)$,

$$G(x) = \frac{1}{Z} \exp(-H(x)/T) \ , \tag{4}$$

in terms of the normalizing partition function $Z$, and the Hamiltonian $H$ operator as the "energy" function,

$$Z = \sum_x \exp(-H(x)/T) \ . \tag{5}$$

For such distributions of states and acceptance probabilities defined by functions such as $h(\Delta E)$, the equilibrium principle of detailed balance holds. I.e., the distributions of states before, $G(x_k)$, and after, $G(x_{k+1})$, applying the acceptance criteria, $h(\Delta E) = h(E_{k+1} - E_k)$ are the same:

$$G(x_k)h(\Delta E(x)) = G(x_{k+1}) \ . \tag{6}$$

This is sufficient to establish that all states of the system can be sampled, in theory. However, the annealing schedule interrupts equilibrium every time the temperature is changed, and so, at best, this must be done carefully and gradually.

An important aspect of the SA algorithm is to pick the ranges of the parameters to be searched. In practice, computation of continuous systems requires some discretization, so without loss of much generality for applications described here, the space can be assumed to be discretized. There are additional constraints that are required when dealing with generating and cost functions with integral values. Many practitioners use novel techniques to narrow the range as the search progresses. For example, based on functional forms derived for many physical systems belonging to the class of Gaussian-Markovian systems, one could choose an algorithm for $g$,

$$g(\Delta x) = (2\pi T)^{-D/2} \exp[-\Delta x^2/(2T)] \,, \tag{7}$$

where $\Delta x = x - x_0$ is the deviation of $x$ from $x_0$ (usually taken to be the just-previously chosen point), proportional to a "momentum" variable, and where $T$ is a measure of the fluctuations of the Boltzmann distribution $g$ in the $D$-dimensional $x$-space. Given $g(\Delta x)$, it has been proven (Geman and Geman, 1984) that it suffices to obtain a global minimum of $E(x)$ if $T$ is selected to be not faster than

$$T(k) = \frac{T_0}{\ln k} \,, \tag{8}$$

with $T_0$ "large enough."

For the purposes of this paper, a heuristic demonstration follows, to show that Eq. (8) will suffice to give a global minimum of $E(x)$ (Szu and Hartley, 1987). In order to statistically assure, i.e., requiring many trials, that any point in $x$-space can be sampled infinitely often in annealing-time (IOT), it suffices to prove that the products of probabilities of not generating a state $x$ IOT for all annealing-times successive to $k_0$ yield zero,

$$\prod_k (1 - g_k) \stackrel{k \to \infty}{=} 0 \,. \tag{9}$$

This is equivalent to

$$\sum_k g_k \stackrel{k \to \infty}{=} \infty \,, \tag{10}$$

as seen by taking the log of Eq. (9) and Taylor expanding in $g_k$. The problem then reduces to finding $T(k)$ to satisfy Eq. (10). Note that, given a very large space to sample, often at best only a "weak" ergodicity can be assumed for this proof, and any such ergodicity even for well-understood physical systems is an open area of research (Ma, 1985).

For BA, if $T(k)$ is selected to be Eq. (8), then Eq. (7) gives

$$\sum_{k=k_0}^{\infty} g_k \geq \sum_{k=k_0}^{\infty} \exp(-\ln k) = \sum_{k=k_0}^{\infty} 1/k = \infty \ . \tag{11}$$

Although there are sound physical principles underlying the choices of Eqs. (7) and (1) (Metropolis *et al*, 1953), it was noted that this method of finding the global minimum in *x*-space was not limited to physics examples requiring *bona fide* "temperatures" and "energies." Rather, this methodology can be readily extended to any problem for which a reasonable probability density $h(\Delta x)$ can be formulated (Kirkpatrick *et al*, 1983).

### 2.1.2. Simulated quenching (SQ)

Many researchers have found it very attractive to take advantage of the ease of coding and implementing SA, utilizing its ability to handle quite complex cost functions and constraints. However, the long time of execution of standard Boltzmann-type SA has many times driven these projects to utilize a temperature schedule too fast to satisfy the sufficiency conditions required to establish a true (weak) ergodic search. A logarithmic temperature schedule is consistent with the Boltzmann algorithm, e.g., the temperature schedule is taken to be

$$T_k = T_0 \frac{\ln k_0}{\ln k} \ , \tag{12}$$

where $T$ is the "temperature," $k$ is the "time" index of annealing, and $k_0$ is some starting index. This can be written for large $k$ as

$$\Delta T = -T_0 \frac{\ln k_0 \Delta k}{k(\ln k)^2} \ , \ k \gg 1$$

$$T_{k+1} = T_k - T_0 \frac{\ln k_0}{k(\ln k)^2} \ . \tag{13}$$

However, some researchers using the Boltzmann algorithm use exponential schedules, e.g.,

$$T_{k+1} = cT_k \ , \ 0 < c < 1$$

$$\frac{\Delta T}{T_k} = (c-1)\Delta k \ , \ k \gg 1$$

$$T_k = T_0 \exp((c-1)k) \ , \tag{14}$$

with expediency the only reason given. While perhaps someday less stringent necessary conditions may be developed for the Boltzmann algorithm, this is not now the state of affairs. The question arises, what is the value of this clear misuse of the claim to use SA to help solve these problems/systems (Ingber, 1993b)? Below, a variant of SA, adaptive simulated annealing (ASA) (Ingber, 1989; Ingber, 1993a), in fact does justify an exponential annealing schedule, but only if a particular distribution is used for the generating function.

### 2.1.3. Fast annealing (FA)

Although there are many variants and improvements made on the "standard" Boltzmann algorithm described above, many textbooks finish just about at this point without going into more detail about other algorithms that depart from this explicit algorithm (van Laarhoven and Aarts, 1987). Specifically, it was noted that the Cauchy distribution has some definite advantages over the Boltzmann form (Szu and Hartley, 1987). The Cauchy distribution they define is

$$g(\Delta x) = \frac{T}{(\Delta x^2 + T^2)^{(D+1)/2}} \ , \tag{15}$$

which has a "fatter" tail than the Gaussian form of the Boltzmann distribution, permitting easier access to test local minima in the search for the desired global minimum.

It is instructive to note the similar corresponding heuristic demonstration, that the Cauchy $g(\Delta x)$ statistically finds a global minimum. If Eq. (8) is replaced by

$$T(k) = \frac{T_0}{k} \ , \tag{16}$$

then here

$$\sum_{k_0}^{\infty} g_k \approx \frac{T_0}{\Delta x^{D+1}} \sum_{k_0}^{\infty} \frac{1}{k} = \infty \ . \tag{17}$$

Note that the "normalization" of $g$ has introduced the annealing-time index $k$, giving some insights into how to construct other annealing distributions. The method of FA is thus seen to have an annealing schedule exponentially faster than the method of BA. This method has been tested in a variety of problems (Szu and Hartley, 1987).

## 2.2. Adaptive Simulated Annealing (ASA)

In a variety of physical problems we have a $D$-dimensional parameter-space. Different parameters have different finite ranges, fixed by physical considerations, and different annealing-time-dependent sensitivities, measured by the curvature of the cost-function at local minima. BA and FA have $g$ distributions which sample infinite ranges, and there is no provision for considering differences in each parameter-dimension, e.g., different sensitivities might require different annealing schedules. These are among several considerations that gave rise to Adaptive Simulated Annealing (ASA). Full details are available by obtaining the publicly available source code (Ingber, 1993a).

ASA considers a parameter $\alpha_k^i$ in dimension $i$ generated at annealing-time $k$ with the range

$$\alpha_k^i \in [A_i, B_i] \,, \tag{18}$$

calculated with the random variable $y^i$,

$$\alpha_{k+1}^i = \alpha_k^i + y^i (B_i - A_i) \,,$$

$$y^i \in [-1, 1] \,. \tag{19}$$

Define the generating function

$$g_T(y) = \prod_{i=1}^{D} \frac{1}{2(|y^i| + T_i) \ln(1 + 1/T_i)} \equiv \prod_{i=1}^{D} g_T^i(y^i) \,, \tag{20}$$

where the subscript $i$ on $T_i$ specifies the parameter index, and the $k$-dependence in $T_i(k)$ for the annealing schedule has been dropped for brevity. Its cumulative probability distribution is

$$G_T(y) = \int_{-1}^{y^1} \cdots \int_{-1}^{y^D} dy'^1 \cdots dy'^D \, g_T(y') \equiv \prod_{i=1}^{D} G_T^i(y^i) \,,$$

$$G_T^i(y^i) = \frac{1}{2} + \frac{\operatorname{sgn}(y^i)}{2} \frac{\ln(1 + |y^i|/T_i)}{\ln(1 + 1/T_i)} \,. \tag{21}$$

$y^i$ is generated from a $u^i$ from the uniform distribution

$$u^i \in U[0, 1] \,,$$

$$y^i = \operatorname{sgn}(u^i - \frac{1}{2}) T_i [(1 + 1/T_i)^{|2u^i - 1|} - 1] \,. \tag{22}$$

It is straightforward to calculate that for an annealing schedule for $T_i$

$$T_i(k) = T_{0i} \exp(-c_i k^{1/D}) \,, \tag{23}$$

a global minima statistically can be obtained. I.e.,

$$\sum_{k_0}^{\infty} g_k \approx \sum_{k_0}^{\infty} [\prod_{i=1}^{D} \frac{1}{2|y^i|c_i}] \frac{1}{k} = \infty \,. \tag{24}$$

It seems sensible to choose control over $c_i$, such that

$$T_{fi} = T_{0i} \exp(-m_i) \text{ when } k_f = \exp n_i \,,$$

$$c_i = m_i \exp(-n_i/D) \,, \tag{25}$$

where $m_i$ and $n_i$ can be considered "free" parameters to help tune ASA for specific problems.

It has proven fruitful to use the same type of annealing schedule for the acceptance function $h$ as used for the generating function $g$, i.e., Equations (23) and (25), but with the number of acceptance points, instead of the number of generated points, used to determine the $k$ for the acceptance temperature.

In one implementation of this algorithm, new parameters $\alpha_{k+1}^i$ are generated from old parameters $\alpha_k^i$ by generating the $y^i$'s until a set of $D$ are obtained satisfying the range constraints. In another alternative supported in ASA, useful for some constraint problems, the $y^i$'s are generated sequentially for each test of the cost function.

### 2.2.1. Reannealing

Whenever doing a multi-dimensional search in the course of a real-world nonlinear physical problem, inevitably one must deal with different changing sensitivities of the $\alpha^i$ in the search. At any given annealing-time, it seems sensible to attempt to "stretch out" the range over which the relatively insensitive parameters are being searched, relative to the ranges of the more sensitive parameters.

It has proven fruitful to accomplish this by periodically rescaling the annealing-time $k$, essentially reannealing, every hundred or so acceptance-events (or at some user-defined modulus of the number of accepted or generated states), in terms of the sensitivities $s_i$ calculated at the most current minimum value of the cost function, $\underline{L}$,

$$s_i = \partial \underline{L}/\partial \alpha^i \,. \tag{26}$$

In terms of the largest $s_i = s_{\max}$, a default rescaling is performed for each $k_i$ of each parameter dimension, whereby a new index $k'_i$ is calculated from each $k_i$,

$$k_i \rightarrow k'_i ,$$

$$T'_{ik'} = T_{ik}(s_{\max}/s_i) ,$$

$$k'_i = (\ln(T_{i0}/T_{ik'})/c_i)^D . \tag{27}$$

$T_{i0}$ is set to unity to begin the search, which is ample to span each parameter dimension.

The acceptance temperature is similarly rescaled. Since the initial acceptance temperature is set equal to an initial trial value of $\underline{L}$, this is typically very large relative to the current best minimum, which may tend to distort the scale of the region currently being sampled. Therefore, when this rescaling is performed, the initial acceptance temperature is reset to the maximum of the most current minimum and the best current minimum of $\underline{L}$, and the annealing-time index associated with this temperature is reset to give a new temperature equal to the minimum of the current cost-function and the absolute values of the current best and last minima.

Also generated are the "standard deviations" of the theoretical forms, calculated as $[\partial^2 \underline{L}/(\partial \alpha^i)^2]^{-1/2}$, for each parameter $\alpha_i$. This gives an estimate of the "noise" that accompanies fits to stochastic data or functions. At the end of the run, the off-diagonal elements of the "covariance matrix" are calculated for all parameters. This inverse curvature of the theoretical cost function can provide a quantitative assessment of the relative sensitivity of parameters to statistical errors in fits to stochastic systems.

### 2.2.2. Quenching

Another adaptive feature of ASA is its ability to perform quenching in a methodical fashion. This is applied by noting that the temperature schedule above can be redefined as

$$T_i(k_i) = T_{0i} \exp(-c_i k_i^{Q_i/D}) ,$$

$$c_i = m_i \exp(-n_i Q_i/D) , \tag{28}$$

in terms of the "quenching factor" $Q_i$. The above proof fails at Eq. (24) if $Q_i > 1$ as

$$\sum_k \prod_k^D 1/k^{Q_i/D} = \sum_k 1/k^{Q_i} < \infty \ . \tag{29}$$

This simple calculation shows how the "curse of dimensionality" arises, and also gives a possible way of living with this disease. In ASA, the influence of large dimensions becomes clearly focussed on the exponential of the power of $k$ being $1/D$, as the annealing required to properly sample the space becomes prohibitively slow. So, if we cannot commit resources to properly sample the space ergodically, then for some systems perhaps the next best procedure would be to turn on quenching, whereby $Q_i$ can become on the order of the size of number of dimensions.

The scale of the power of $1/D$ temperature schedule used for the acceptance function can be altered in a similar fashion. However, this does not affect the annealing proof of ASA, and so this may used without damaging the (weak) ergodicity property.

### 2.2.3. ASA applications

The above defines this method of adaptive simulated annealing (ASA), previously called very fast simulated reannealing (VFSR) (Ingber, 1989) only named such to contrast it the previous method of fast annealing (FA) (Szu and Hartley, 1987). The annealing schedules for the temperatures $T_i$ decrease exponentially in annealing-time $k$, i.e., $T_i = T_{i0} \exp(-c_i k^{1/D})$. Of course, the fatter the tail of the generating function, the smaller the ratio of acceptance to generated points in the fit. However, in practice, it is found that for a given generating function, this ratio is approximately constant as the fit finds a global minimum. Therefore, for a large parameter space, the efficiency of the fit is determined by the annealing schedule of the generating function.

A major difference between ASA and BA algorithms is that the ergodic sampling takes place in an $n + 1$ dimensional space, i.e., in terms of $n$ parameters and the cost function. In ASA the exponential annealing schedules permit resources to be spent adaptively on reannealing and on pacing the convergence in all dimensions, ensuring ample global searching in the first phases of search and ample quick convergence in the final phases.

I have used ASA in several systems, ranging from combat analysis (Ingber, Fujio, and Wehner, 1991; Ingber and Sworder, 1991), to finance (Ingber, 1990; Ingber, Wehner *et al*, 1991), to neuroscience (Ingber, 1991), to a set of test problems (Ingber and Rosen, 1992), to a new technique combining the power of SA with the physics of large-scale systems (Ingber, 1992).

An optimization algorithm typically is just one tool used in a major project, and many users of ASA use this code as just one such tool. Furthermore, the author has made it a policy not to give out any information he receives from users unless they specifically permit him to do so. Thus, many people ask questions and give feedback on the code they might not give otherwise. In order to get maximum feedback without unduly bothering researchers, the e-mail ASA_list is not open forum, but rather an efficient moderated medium to gather information. Some published acknowledgments to use of the code are in the asa_papers file of the ASA archive; the disciplines range from physics (Brown *et al*, 1994; Tang *et al*, 1995) to neural networks (Cohen, 1994; Indiveri *et al*, 1993) to difficult imaging problems (Wu and Levine, 1993) to finance (Wofsey, 1993), where most of the latter uses are kept proprietary.

## 3. ASA OPTIONS

### 3.1. Options

ASA likely is the most powerful and flexible SA code presently available, because the code has benefited from the feedback of the many users, and their feedback has been used to add much to the code beyond the basic ASA algorithm described above.

The code has two basic modules in the ASA C-code, a user and an asa module. All OPTIONS in the code have been tested to work with templates provided in the user module. Feedback has developed a code which seems to run well across many platforms, e.g., PC's, Macs, Crays, many UNIX workstations, etc.

The emphasis in development of ASA has been to add power and flexibility wherever possible. To make these extra features and code accessible to non-expert programmers, a "meta-language" of OPTIONS is used. Many of these OPTIONS can be set in the provided Makefile, an asa_opt data file from which to read in information, arguments passed to the compilation procedures, or in the user module files (user.c, user.h , and asa_user.h, the latter being a bridge between the asa and user modules).

A price is paid for this continual development of a more powerful and more flexible code. The new user is presented with many OPTIONS, on the order of a hundred. In many cases, when the ASA default OPTIONS work fine, only the user's own call to his/her cost function is required. However, if these defaults are not suitable for a particular system, then the user can become bewildered by the many OPTIONS. If not much is known *a priorí* about the system to be optimized, then the task is to try to find the values of the OPTIONS appropriate to the given system. The less known about the system, the harder is this task.

Experiences support the premise that the output of the code, using the ASA_PRINT_MORE OPTIONS to give information at each new best accepted state, often can be used to diagnose problems in annealing. Eventually, I hope that enough experience will be generated, to be able to develop some kind of graphical menu-driven expert system to help guide users to optimize a wide range of cost functions. Beyond this, I dream of a day when such a user-friendly shell can guide users to alternative optimization algorithms at different scales and stages of the optimization process.

### 3.2. Examples of OPTIONS

The following discussion of some of the OPTIONS available in ASA also serves to illustrate the typical kinds of problems many users have with their particular systems, and some of the approaches that SA can offer to face these problems. The OPTIONS are organized into three group. The DEFINE_OPTIONS comprise two set of OPTIONS, the Pre-Compile DEFINE_OPTIONS and the Printing DEFINE_OPTIONS, which are called at the time of compilation; these comprise about half of the OPTIONS. The other Program OPTIONS are housed in a structure passed with the cost function, and together with the other parameters passed in the cost function, these can be modified adaptively. That is, they can be changed within the cost function to take effect upon reentering the asa program.

#### 3.2.1. Integer and continuous parameters

ASA can accommodate mixtures of integer and continuous parameters. This is accomplished quite simply, with a small overhead for integers, by truncating generated floating-point numbers within sensible integral windows. There have been many "rumors" that SA can only handle integer or continuous parameters, but these statements are unsupported.

### 3.2.2. Constraints

One of the immediate attractions of SA to people trying to optimize complex systems is the ease with which SA can accommodate complex constraints. Typically, there is no need for penalty functions, etc. Generated points that do not satisfy the constraints are simply rejected before trying any acceptance test.

Equality constraints, if processed as above, present a problem for any global optimization that relies on sampling, because the search is being constrained on the surface of some volume, and the entire volume is being sampled. Therefore, it is recommended that the user first numerically substitute solution(s) of the the equalities for some parameters. For example, if the cost function $C$ has $n$ parameters, $C(p_1, p_2, \cdots, p_n)$, and an equality constraint exists between parameters $p_n$ and $p_{n-1}$, then solve this equation for $p_n$, numerically or algebraically, redefining the cost function to one with $n-1$ parameters, $C'$. If the solution to this equation, or perhaps a set of $m$ such equality constraints to reduce the number of parameters actually processed by ASA to $n-m$, is not simply written down, then such constraints must be solved with other algorithms within the cost function.

### 3.2.3. Annealing scales

Perhaps the easiest to understand problem that can arise when using SA, also is the the most often neglected. I get many queries that start out questioning why ASA doesn't immediately find the global optimal point? The answer most often lies in the scaling parameters used in annealing the parameter and/or cost temperatures.

For example, if the search is carried out in a system with several local minima, but the temperature is too low so that only rarely can the search sample these minima, it may take an extremely long time with arbitrarily good numerical precision to eventually sample these minima as normal annealing proceeds to lower and lower temperatures. Clearly, it would be better to have the starting temperature at the scale in question be commensurately larger, and perhaps be cooled more slowly.

(a) *Parameter temperatures*. In some SA algorithms, like BA, the starting temperature controls the values of temperature encountered at subsequent stages of search. In ASA, because of the finite ranges of the parameters, the parameter temperatures are started to establish to a fat tail throughout the range; the exponential annealing rates usually permit selecting even quite large initial ranges to be sure of covering all optima. There are free ASA-parameters for each temperature to scale its exponential decrease, without affecting the basic sampling proof.

(b) *Cost temperature*. The annealing scale for the cost temperature, also called the acceptance temperature, affects the rate of narrowing the window of the Boltzmann acceptance test. In ASA, this scale can be adaptively changed, and even the Boltzmann test can be changed to a different distribution.

### 3.2.4. Reannealing

(a) *Parameter temperatures*. For the parameter temperatures, the tangents (or any other alternative functions that might be defined by the user) are used as a relative measure of the "steepness" of each dimension the most recent best saved state. As demonstrated for the ASA_TEST problem (Ingber, 1993b), this feature can enhance the efficiency of the search.

(b) *Cost temperature*. For the cost temperature, a separate OPTIONS permits rescaling of the cost temperature to be set to the scale of the minimum of the current cost temperature and the absolute values of the last and best saved minima, to keep the acceptance test sensitive at a reasonable scale. This can be extremely important of the system's terrain changes with the scale of the search. This procedure also may need to radically altered, possible with other OPTIONS, if the search early becomes stuck in local optima, e.g., because the system's terrain abruptly changes with the scale of the search.

### 3.2.5. Quenching

An SA algorithm loses much of its authority if the search "cheats" by trying to anneal at rates faster than permitted by its associated proof, e.g., simulated quenching (SQ). However, this can be useful in a number of circumstances (Ingber, 1993b).

When the dimension of a parameter space, each parameter having a continuous or large integral set of values, reaches 15 to 20, the volume of search typically becomes quite large and this can severely tax most present-day workstations. Instead of just giving up on SA and trying a different "greedy" and/or quasi-Newton algorithm, ASA provides a methodical way to deviate from SA into SQ algorithms.

As another use of quenching, one that does not necessarily violate any sampling proof, it may be useful in the course of search to adaptively drop subsets of parameters that seem to have been reasonably optimized relative to other parameters. The remaining parameters can then be more efficiently searched within their smaller dimensional space, by adjusting the dependence of the annealing to the new dimension. This can be accomplished conveniently with the QUENCHing OPTIONS.

### 3.2.6. ASA sampling

Since ASA accomplishes its fit by importance sampling the space of parameters, it would seem that this process should provide a good sampling technique for other purposes, e.g., performing integrals. Of course, as stated above, the use of Monte Carlo techniques for performing integrals (Metropolis *et al*, 1953) is generally credited to be the origin of the development of SA (Kirkpatrick *et al*, 1983). However, importance sampling with the fastest permitted temperature schedules often can lead to quite poor resolutions of local minima which may substantially contribute to integrals. Then, the rates of annealing must be slowed down, e.g., using inverse QUENCHing, to get better resolution. The ASA_SAMPLE OPTIONS collects the generating and acceptance biases incurred during importance sampling, so that this information can be used more generally than for just finding the optimal point of the fit.

### 3.2.7. Self optimization

An advantage of C code over some other languages is the relative ease by which recursive calls can be implemented. Some care must be taken to keep variables local to each subroutine. In its current form ASA can recursively call itself. Some complex problems, possessing nests of optimized systems, require this.

If not much information is known about a particular system, if the ASA defaults do not seem to work very well, and if after a bit of experimentation it still is not clear how to select values for some of the ASA OPTIONS, then the SELF_OPTIMIZE OPTIONS can be very useful. This sets up a top level search on the ASA OPTIONS themselves, using criteria of the system as its own cost function, e.g., the best attained optimal value of the system's cost function (the cost function for the actual problem to be solved) for each given set of top level OPTIONS, or the number of generated states required to reach a given value of the system's cost function, etc. Since this can consume a lot of CPU resources, it is recommended that only a few ASA OPTIONS and a scaled down system cost function or system data be selected for this OPTIONS.

Even if good results are being attained by ASA, SELF_OPTIMIZE can be used to find a more efficient set of ASA OPTIONS. I think that this kind of OPTIONS would be useful for most useful for many non-linear optimization algorithms. Many of the OPTIONS broken out in clear view in ASA are similarly represented but "hidden" within the code of other algorithms. Self optimization of such parameters can be very useful for production runs of complex systems.

### 3.2.8. Alternative distributions/functions

There are OPTIONS to permit replacing or modifying the functions and distributions used in the asa module. Foe example, modifications can be made of the generating function (e.g., variants of the Boltzmann and Cauchy distributions are given in the user module), the acceptance function (e.g., a class of functions that asymptotically approach the Boltzmann function is given in the user module), and the reannealing functions used to rescale the parameter and cost temperatures.

### 3.2.9. Parallel code

It is quite difficult to directly parallelize an SA algorithm (Ingber, 1993b), e.g., without incurring very restrictive constraints on temperature schedules (Kimura and Taki, 1991), or violating an associated sampling proof (Frost, 1993). However, the fat tail of ASA permits parallelization of developing generated states prior to subjecting them to the acceptance test (Ingber, 1992). The ASA_PARALLEL OPTIONS provide parameters to easily parallelize the code, using various implementations, e.g., PVM, shared memory, etc.

The scale of parallelization afforded by ASA, without violating its sampling proof, is given by a typical ratio of the number of generated to accepted states. Several experts in parallelization suggest that massive parallelization e.g., on the order of the human brain, may take place quite far into the future, that this might be somewhat less useful for many applications than previously thought, and that most useful scales of parallelization might be on scales of order 10 to 1000. Depending on the specific problem, such scales are common in ASA optimization, and the current ASA code can implement such parallelization.

No specific parallel implementation has yet been included in the code. A project was set up under an National Science Foundation (NSF) grant of Cray time at the Pittsburgh Supercomputer Center for this purpose, but we soon realized that the grant rules restricted us to first vectorize the ASA code on a C90.

This seemed like a lot of effort for a specific architecture, e.g., an effort that would not even translate into useful code on a T3D, and yet this still would not address the true parallelization of the ASA code. Therefore, our volunteer group went as far as to confirm the nature of the required code, and I placed the above "hooks" into the present code. People who played a major role in examining these issues were Tim Burns of the Utah Supercomputing Institute, Alan Cabrera of Sanwa Financial Products Co., and Wolfram Gloger of Poliklinik für Zahnerhaltung und Parodontologie.

## 4. IV. MISLEADING SA CLAIMS

### 4.1. General Considerations

#### 4.1.1. Comparisons among algorithms

There is not much doubt that, for any reasonably difficult nonlinear or stochastic system, some optimization codes will perform better than others. Nonlinear systems are typically non-typical, and so it should be expected that some algorithms are better suited for some systems than for others. It is unlikely that there ever will be developed a "black-box canned" algorithm as exist for many linear systems.

Thus, comparisons among algorithms on simple "toy" problems, while they may be interesting and necessary (and therefore are included below), perhaps exposing something of the nature of the tested algorithms, may offer little insight to help a given researcher faced with a new problem. Most likely, the researcher must work to learn more about the system to best apply a given algorithm.

There is a formal proof that all algorithms should perform the same over all problems (Wolpert and Macready, 1995). This obviously is of little practical help for a given particular problem, albeit the authors claim that their approach offers some guidelines for specific algorithms. For example, it would be ill-advised to draw the conclusion that therefore one might as well always just use Newton's algorithm on all problems faced by a research group. However, such research is necessary to bring more objectivity to the present "art" of global optimization.

#### 4.1.2. Tuning algorithms for specific problems

There is not much doubt that, for any reasonably difficult nonlinear or stochastic system, a given optimization algorithm can be "tuned" to enhance its performance. Indeed, since it takes some time and effort to become familiar with a given code, the ability to tune a given algorithm for use in more than one problem should be considered an important feature of an algorithm.

### 4.2. Examples of Comparisons

#### 4.2.1. Genetic algorithms (GA)

A direct comparison was made between ASA/VFSR and a publicly available genetic algorithm (GA) code, using a test suite already adapted and adopted for GA (Schraudolph and Grefenstette, 1991). In each case, ASA outperformed the GA problem.

It should be clear that GA is a class of algorithms that are interesting in their own right. GA was not originally developed as an optimization algorithm (De Jong, 1992), and basic GA does not offer any statistical guarantee of global convergence to an optimal point (Forrest, 1993). Nevertheless, it should be expected that GA may be better suited for some problems than SA.

#### 4.2.2. Comparing BA, FA, and ASA

A study was made comparing the performance among Boltzmann annealing (BA), fast annealing (FA), and ASA, using the difficult test problem that comes with the ASA code (Rosen, 1992). Only ASA regularly attained the global minimum, and it was more efficient in attaining regular minima at each comparable number of generated states than BA or FA. The ASA_TEST problem contains $10^{20}$ local minima with a parameter dimension of 4 (Corana *et al*, 1987), and it typically takes 2000–3000 generated states, or about 2 CPU-sec on a Sun SPARC-II, to find the global minimum.

#### 4.2.3. Dynamic hill climbing (DHC)

Michael de la Maza posted notices to public electronic bulletin boards, e.g., as summarized in a public mailing list GA-List@AIC.NRL.NAVY.MIL, that his new algorithm, dynamic hill climbing (DHC), clearly outperformed genetic algorithms and ASA. His code is available by sending e-mail to

dhc@ai.mit.edu. Since DHC is a variant of a "greedy" algorithm, it seemed appropriate to permit ASA to also enter its quenching (SQ) domain. The following excerpt is the reply posting in the above bulletin board.

"SQ techniques like GA obviously are important and are crucial to solving many systems in time periods much shorter than might be obtained by SA. In ASA, if annealing is forsaken, and QUENCHing turned on, voiding the proof of sampling, remarkable increases of speed can be obtained, apparently sometimes even greater than other 'greedy' algorithms. For example, in (Ingber and Rosen, 1992) along with 5 GA test problems from the UCSD GA archive, another harder problem (the ASA_TEST problem that comes with the ASA code) was used. In (Ingber, 1993b) QUENCHing was applied to this harder problem. The resulting SQ code was shown to speed up the search by as much as as factor of 86 (without even attempting to see if this could be increased further with more extreme quenching). This is greater than the factor of 30 that was reported by Michael de la Maza for Dynamic Hill Climbing (DHC). This is a simple change of one number in the code, turning it into a variant of SQ, and is not equivalent to 'tuning' any of the other many ASA options, e.g., like SELF_OPTIMIZE, USER_COST_SCHEDULE, etc. Note that SQ will not suffice for all systems; several users of ASA reported that QUENCHing did not find the global optimal point that was otherwise be found using the 'correct' ASA algorithm."

### 4.2.4. Tsallis statistics

A recent paper claimed that a statistics whose parameterization permits an asymptotic approximation to the exponential function used for the Boltzmann of the standard SA acceptance test, Tsallis statistics, is superior to the Boltzmann test (Penna, 1994), and an example was given comparing standard SA to this new algorithm in the traveling salesman problem (TSP). There are two issues here: (a) the value of the Tsallis test vs the Boltzmann test, and (b) the use of TSP for the confirmation of (a).

It seems very reasonable that the Tsallis test should be better than the Boltzmann test for the SA acceptance test. For example, if the Boltzmann statistics did well on a given cost function $C$, then it might be the case that for the cost function $C' = \exp(C)$ a more moderate test, such as obtained for some parameterizations of the Tsallis statistics, would be more appropriate to avoid getting stuck in local minima of $C'$. In fact, from its first inception VFSR and ASA have included parameters to effect similar alternatives, and the latest versions of ASA now have the Tsallis statistics as another alternative that can be commented out. I have not yet experienced any advantages of this over the Boltzmann test when other ASA alternatives are permitted to be used, but it seems likely that there do exist some problems that might benefit by its use.

The use of TSP as a test for comparisons among SA techniques seems quite inappropriate. To quote another source (Wolpert and Macready, 1995): "As an example of this, it is well known that generic methods (like simulated annealing and genetic algorithms) are unable to compete with carefully hand-crafted solutions for specific search problems. The Traveling Salesman (TSP) Problem is an excellent example of such a situation; the best search algorithms for the TSP problem are hand-tailored for it (Reinelt, 1994)."

### 4.2.5. Shubert problem

In (Cvijović and Klinowski, 1995) some strong claims about the superiority of taboo search over genetic algorithms and simulated annealing were made.

I took the only function given in that paper that also gave all parameters in this article, the Shubert function, and ran a straight ASA test with absolutely no tuning. (I do consider tuning essential for optimum use of any global optimization algorithm.) Table 3 of that paper states that the results were obtained by counting the number of generated cost functions needed to first attain a known global minimum of -186.7309. (There are 18 global minima out of 760 local minima). That paper gives results for taboo search (TS), and two variants of simulated annealing, SA1 and SA2, averaged over 100 runs. SA1 required 241,215 generated cost functions; SA2 required 780 generated cost functions; TS required 727 generated cost functions. The same test was performed with ASA, and it was found that ASA required 577 generated cost functions. More details of the coding of this problem and the results of the ASA calculation are given in the NOTES file of the ASA code.

The result of course will vary somewhat with the initial seed for the random number generator and with the initial guess for the 2 parameters. As argued for many other cases, the fat tail of ASA, creating a rather large generated to acceptance ratio, usually insures that the variance of the results is not too large. (Note that here there are 18 global minima which may cause more of a spread.) The 100 iterations of ASA used the default to let ASA randomly select the initial values for the annealing loop (as was implied for SA1, SA2, and TS). The results from ASA (similar results not reported in that paper) were a minimum number of generated point to be 231, a maximum number of generated point to be 1352, a mean of 577.1, and a standard deviation of 145.6.

### 4.2.6. Colville problems

In early 1993 I was contacted by Zbigniew Michalewicz, author of a new code, GENECOP, a GA-based optimization package available via anonymous ftp from unccsun.uncc.edu in directory coe/evol. He stated that the ASA code failed to get any reasonable fit to two Colville problems, a set of problems developed by Colville, Floudas and Pardalos. He demonstrated this by offering his own results with GENECOP in which he obtained better results that that obtained previously in the literature on one these problems.

On one problem, after I learned that GENECOP first made algebraic substitutions to eliminate equality constraints, this also was performed this before using ASA/VFSR. GENECOP had produced better results than previously obtained in the literature for this example, and ASA/VFSR obtained even better results than GENECOP. The previously published results gave a final cost of -47.707579; GENECOP gave cost of -47.760765; ASA/VFSR gave cost of -47.76109. For such systems, I have suggested that GENECOP be used as a front end to ASA.

On the other problem, using the default parameters of ASA, the code did indeed hover in a local minimum, likely for so long that it might take an enormous amount of human time and unrealistic computer precision (to keep sampling at incredibly small temperatures) to find the global minimum. However, just a few minutes of modestly tuning a few ASA parameters produced the global minimum very efficiently.

More details of the coding of these two problems and the results of the ASA calculations are given in the NOTES file of the ASA code.

### 5. CONCLUSION

If asked to state one major common feature of nonlinear systems, in the context of optimization, the feature most likely should be given is that nonlinear systems typically are non-typical. It is unlikely that any "canned black-box" code can be developed, requiring no or few minor adjustments, that will usefully guarantee efficient global optimization for severely nonlinear systems, e.g., similar to what might be expected for many quasi-linear systems.

In the absence of knowledge about a particular system, given that only SA can offer at least a "statistical" proof of global optimization, then the first algorithm of choice clearly is SA. Modifications of SA, e.g., SQ quenching algorithms, may be competitive with other techniques, e.g., simplex or genetic algorithms, but among these the best choice is not so clear. SQ does offer a relatively simple approach to quickly writing code for optimization, but ultimately the end results must justify this means.

This argument for the use of SA has an opposite side. If some information about a system can be incorporated into some other global optimization technique, and it can be determined that the technique can deliver the global optimum point, often that technique will be more efficient than SA. E.g., a quasi-Newton algorithm will be more efficient than SA for parabolic systems.

For many researchers, the first choice of algorithm to use for a nonlinear or stochastic problem likely will be one with which they already are familiar. If that fails, then SA is an option to try next. More research needs to be done to see if a more objective classification of nonlinear systems can be developed to help guide a given researcher to a given algorithm for a given problem. As the examples included in the documentation of the ASA code illustrate, there have been "surprises" whereby some very difficult problems have been quickly solved by ASA, while others have required quite a bit of "tuning" to establish a good set of starting OPTIONS.

Especially among first-time users of SA, often there is much misunderstanding and lack of appreciation of just what an SA code can immediately do for a particular problem. Some education is necessary to make users aware of the potential problems that may arise and what remedies the particular algorithm can offer to overcome these obstacles.